# MICA: A fast short-read aligner that takes full advantage of Intel® Many Integrated Core Architecture (MIC)

Sze-Hang Chan[1,†], Jeanno Cheung[1,†], Edward Wu[1,†], Heng Wang[1,2,†], Chi-Man Liu[1], Xiaoqian Zhu[2], Shaoliang Peng[2], Ruibang Luo[1,*], Tak-Wah Lam[1,*]

[1] HKU-BGI Bioinformatics Algorithms and Core Technology Research Laboratory, University of Hong Kong, Hong Kong. [2] School of Computer Science, National University of Defense Technology, China. [†]Joint first authors.

**ABSTRACT**
**Background:** Short-read aligners have recently gained a lot of speed by exploiting the massive parallelism of GPU. An uprising alternative to GPU is Intel MIC; supercomputers like Tianhe-2, currently top of TOP500, is built with 48,000 MIC boards to offer ~55 PFLOPS. The CPU-like architecture of MIC allows CPU-based software to be parallelized easily; however, the performance is often inferior to GPU counterparts as an MIC board contains only ~60 cores (while a GPU board typically has over a thousand cores).
**Results:** To better utilize MIC-enabled computers for NGS data analysis, we developed a new short-read aligner MICA that is optimized in view of MIC's limitation and the extra parallelism inside each MIC core. Experiments on aligning 150bp paired-end reads show that MICA using one MIC board is 4.9 times faster than the BWA-MEM (using 6-core of a top-end CPU), and slightly faster than SOAP3-dp (using a GPU). Furthermore, MICA's simplicity allows very efficient scale-up when multiple MIC boards are used in a node (3 cards give a 14.1-fold speedup over BWA-MEM).
**Summary:** MICA can be readily used by MIC-enabled supercomputers for production purpose. We have tested MICA on Tianhe-2 with 90 WGS samples (17.47 Tera-bases), which can be aligned in an hour less than 400 nodes. MICA has impressive performance even though the current MIC is at its initial stage of development (the next generation of MIC has been announced to release in late 2014).
**Availability and Implementation:** MICA is under BSD (3-Clause) and freely available at http://sourceforge.net/projects/mica-aligner
**Contact:** rbluo@cs.hku.hk, twlam@cs.hku.hk
**Supplementary information:** Supplementary information is available at Bioinformatics online.

## 1 INTRODUCTION

The recently announced Illumina HiSeq X Ten sequencing platform enables the first $1,000 sequencing of human genome. In one year's time, one HiSeq X Ten system is able to sequence 18,000 whole human genomes at 30x coverage, and four such systems are capable of sequencing more genomes than in all of history. With this super increase in sequencing capacity, efficient algorithms that fully utilize the power of acceleration hardware or devices other than CPU is of great essence. For example, SOAP3-dp (Luo, et al., 2013), making use of a graphic processing unit (GPU), is two to tens of times faster than mainstream CPU aligners and delivers higher sensitivity. Besides GPU, Intel's new product called Many Integrated Core (MIC), a.k.a. Xeon Phi Co-processor, has attracted attention; it was introduced in 2011 based on x86 architecture. The latest product line of MIC gives 57-61 cores and 8 GB of memory in one board, providing ~1TFlops. Two top-ten supercomputers in Top500 list are equipped with MIC (Tianhe-2 with 48,000 MIC boards, and Stampede with 6,400 boards). However, up to date, limited study on short read alignment has been carried out on MIC. This paper shows a new aligner MICA, tailor-made for MIC to fully utilize its computing power.

## 2 METHODS: ARCHITECTURE & PARALLELISM

MICA inputs reads from one or multiple files in FASTA/FASTQ format and outputs alignment results in SAM/BAM format. In a typical setting, MICA is running on a server (host) equipped with one or more MIC boards. MICA runs each MIC using the offload mode (instead of the native mode) so as to exploit the host's memory and better I/O performance. The latter is important when dealing with large volume of sequencing data. For each MIC board, MICA maintains a CPU thread in the host called an MIC-controller, which feeds the MIC with a million of reads each time, and spawns 224 MIC threads (running on 56 cores) to align the reads in parallel. At the end, the alignment results are copied to the MIC-controller for output (Figure 1). Inside each thread, MICA adopts a widely used two-phase approach to align each read. First, it exploits the presence of a BWT index of the reference genome in the MIC's global memory to align with few mismatches; if not successful, it then uses dynamic programming to align with Indels.

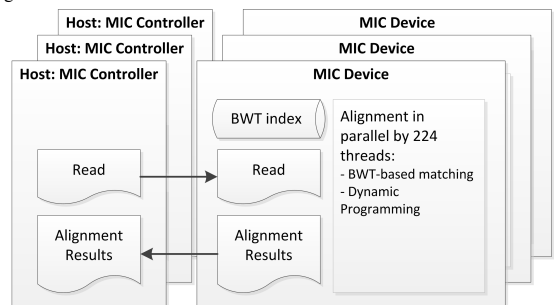

Figure 1. MICA's architecture: interaction between the host and MIC cards.

**Pipelining & load balancing.** To achieve maximum throughput, MICA is designed to always keep the MIC cards busy in aligning reads. Tasks involving delay or waiting (such as I/O) are all handled by the host. Furthermore, each MIC thread occasionally may not align all the reads give to it; this happens when the thread is given too many resource-demanding (repetitive) reads. Unaligned reads will be handled by the CPU side MIC-controller eventually. Such control mechanism is important to enable an effective utilization of all the cores in each MIC.

**Extra parallelism**. We observe that an MIC core is roughly 4 to 6 times slower than a CPU core when running a program originally designed for CPU. This seems to suggest that an MIC with 57 cores is not particularly powerful. Noteworthy, each MIC core comprise thirty-two 512-bit registers, each of which allows sixteen 32-bit data to be operated in parallel. Such extra parallelism, if exploited properly, can boost the efficiency drastically. The design of MICA takes advantage of such extra parallelism in its bottleneck on BWT operations and dynamic programming.

Many efficient aligners are based on BWT indexing. MICA adapts SOAP3-dp's BWT index to work in MIC. This index, designed for excessive number of threads in parallel (SOAP3-dp is GPU-based), has been optimized to reduce memory transaction and avoid memory contention among the threads. MICA has redesigned the operations on the index by taking advantage of 512-bit registers to speed up different arithmetic and memory transaction steps when querying the BWT.

**Highly parallel dynamic programming.** MICA uses dynamic programming (DP) to align reads where indel and soft clipping are necessary. We have implemented DP in MICA in such a way that the parallelization power of 512-bit registers can be fully utilized.

To align a read of length $m$ and a reference region of length $n$ by DP, the traditional approach needs to fill a 2-dimenional table (denoted $T[n,m]$) for storing the intermediate results. $T[i, j]$ is determined by $T[i-1,j-1]$, $T[i-1,j]$ and $T[i, j-1]$ (upper, upper-left and left entry dependency). Such dependency enforces us to compute consecutive entries on a row (or a column) one by one.

In MICA, this table is instead represented by an array, which stores the values in the diagonal order of $T$. E.g., $T[i, j]$ is followed by $T[i-1,j+1]$, $T[i-2,j+2]$ and so on. This representation allows us to fill 16 entries of the DP table in one step. E.g., to fill $T[i, j]$, $T[i-1, j+1]$ … $T[i-15, j+15]$ in a single iteration, we need to load 16 entries from 3 memory locations beginning from $T[i-1, j-i]$, $T[i-1, j]$ and $T[i, j-1]$. In practice, such representation, however, requires rather tedious programming to handle the boundaries correctly.

As noted before, when using MIC, the amount of memory transaction from each core is an important factor when determining the efficiency. To minimize memory transaction, we pack all the necessary information (including the read and reference) required for calculation into a 32-bit entry in the DP table and the memory transactions is reduced to loading 512-bit vectors of the DP table.

We follow the affine gap penalty model, which indeed needs to compute three DP tables $M$, $I$, $D$ to store the optimal score when aligning up to each position with the last character being a Match, Insert or Delete, respectively. To minimize memory transaction, we also pack the $M$, $I$, $D$ values for each $i, j$ into $T[i, j]$.

## 3 EXPERIMENTAL RESULTS

**Comparison with other aligners**. We used 77-fold of 150bp Illumina paired-end reads of the YH samples (PE150, Supplementary) (Luo, et al., 2012) to benchmark MICA and other state-of-the-art aligners including GPU-based SOAP3-dp (Luo, et al., 2013) and CPU-based BWA-MEM (Li, 2013). The results are shown in Table 1; the figures were calculated as the average of three repetitive runs, and a tailor-made setting was used for each aligner to ensure the best performance[1]. When using PE150, MICA using one card is ~4.9 times faster than BWA-MEM and slightly faster than SOAP3-dp. The speed of MICA can be scaled up almost linearly with additional cards. When using three cards, MICA is ~14.1 times faster than BWA-MEM. MICA's sensitivity is 3.1% and 0.47% higher than BWA-MEM and SOAP3-dp, respectively.

Table 1. Performance of MICA, SOAP3-dp and BWA-MEM on real data.

| Dataset | Volume (Gbp) | # of Read Pairs (M) | Fold | MICA | | | | | | SOAP3-dp | | BWA-MEM | |
|---|---|---|---|---|---|---|---|---|---|---|---|---|---|
| | | | | 1 card | 2 cards | 2 cards scale-up | 3 cards | 3 cards scale-up | Properly paired | 1 card | Properly paired | 6 cores | Properly paired |
| PE150 | 232.15 | 773.83 | 77.38 | 20,919s (5.81hr) | 10,618s (2.95hr) | 1.97x | 7,183s (2.00hr) | 2.91x | 95.48% | 25,878s (7.19hr) | 95.01% | 101,466s (28.19hr) | 92.32% |
| PE100 | 148.43 | 742.16 | 49.48 | 15,879s (4.41hr) | 8,093s (2.25hr) | 1.96x | 5,453s (1.51hr) | 2.91x | 97.23% | 17,982s (5.00hr) | 97.08% | 53,832s (14.95hr) | 95.74% |

We have also benchmarked 100bp paired-end reads (Luo, et al., 2012) (PE100, ~49-fold, Supplementary). Interestingly, MICA's acceleration on PE150 is slightly more significant than on PE100, for the latter MICA is ~3.4 and ~9.9 times faster than BWA-MEM when using one and three MIC cards, resp. When processing many shorter reads, MICA has a bottleneck in access to the memory.

**Production test on Tianhe-2 supercomputer**. Tianhe-2 is equipped with 16,000 computing nodes, each with two 6-core CPUs and three MIC cards. To test Tianhe-2 with the workload of a population scale study, we used whole genome sequencing data of 45 CHB and 45 CHS samples from the 1000 genomes project (Supplementary); a sample has 64.68-fold coverage on average. In total, we have 932 lanes of 100bp paired-end reads with size varying from 1.82- to 12.7-fold per lane.

The alignment of 932 lanes was carried out in 3 different settings, making up 2,796 jobs. All jobs were dispatched simultaneously using one node per job. See Table 2 for the results. The mean time for a lane to finish was 1,303 seconds (assuming the SAM format); even the largest lane finished within an hour. The number of nodes can be reduced by packing smaller jobs into one node; hence 400 nodes are sufficient to finish all 932 lanes in an hour.

Table 2. Experiment results on Tianhe-2 supercomputer.

| Setting | MIC card used | Output Format | Finished lane | Longest (sec) | Mean (sec) | Median (sec) |
|---|---|---|---|---|---|---|
| 1 | 3 | SAM | 923 | 3,425 | 1,303 | 1,220 |
| 2 | 3 | BAM | 929 | 6,012 | 2,777 | 2,666 |
| 3 | 3 | 6-thread BAM | 928 | 4,484 | 1,371 | 1,321 |

MICA is slower when outputting in BAM format than in SAM format. This is due to a bottleneck in the CPU rather than in the MIC, which can be circumvented by using more CPU threads.

## ACKNOWLEDGEMENTS

*Funding*: Hong Kong GRF HKU-713512E and ITF GHP/011/12.
*Conflict of Interest*: None declared.

---

[1] MICA: *version v0.1.190. Compiled with Intel C/C++ Compiler version 13.1; one to three 57-core MIC cards (8G, ECC enabled); each coupled with a CPU core; one CPU core for output (in SAM format).*
SOAP3-dp: *version r176. Compiled with GCC 4.7.2 and CUDA SDK 5.5; one nVidia GeForce GTX680 (4G); 4 CPU cores; serial BAM output.*
BWA-MEM: *version 0.7.5a. Compiled with GCC 4.7.2; 6 CPU cores; SAM output.*
Host machine: *Intel i7-3730k, 6-core @3.2GHz; 64G memory.*